\documentclass[a4paper,10pt]{JHEP3}
\usepackage{epsfig}
\usepackage{amsfonts,amsmath,amssymb}
\newcommand{\f}[2]{\frac{#1}{#2}}
\newcommand{\mc}[1]{\mathcal{#1}}

\title{\bf Cosmological dynamics of brane f(R) gravity}
\author{Zahra Haghani
\footnote{z$\_$haghani@sbu.ac.ir}
, Hamid Reza Sepangi
\footnote{hr-sepangi@sbu.ac.ir}
and Shahab Shahidi
\footnote{s$\_$shahidi@sbu.ac.ir} \\
Department of Physics, Shahid Beheshti University, G. C.,
Evin,Tehran 19839, Iran}
\maketitle
\abstract{
The cosmological dynamics of a brane world scenario where the bulk action is taken as a generic function of the Ricci scalar is considered in a framework where the use of the $\mathbb{Z}_2$ symmetry and Israel junction conditions are relaxed. The corresponding cosmological solutions  for some specific forms of $f(\mc{R})$ are obtained and shown to be in the form of exponential as well as power law for a vacuum brane space-time. It is shown that the existence of matter dominated epoch for a bulk action in the form of a power law for $\cal R$ can only be obtained in the presence of ordinary matter. Using  phase space analysis, we show that the universe must start from an unstable matter dominated epoch and eventually falls into a stable accelerated expanding phase.
}
\begin{document}

                              \section{Introduction}\label{sec1}
In 1999, Randall and Sundrum (RS) \cite{randall} proposed a model in
which our $4D$ universe, the brane, is embedded in an AdS$_5$ bulk
space. All gauge fields are confined to the brane while gravity can
propagate into the extra dimension and thus into the bulk space. As
is now well-known, such a proposal opened a new window through which
a fresh look at the universe has become a possibility. One outcome
of the RS model was that it offered a rational explanation to the
question of hierarchy; the enormous disparity between the
fundamental forces. Since its appearance, numerous scenarios based
on the RS setup have been proposed, each dealing with some
particular aspects of the structure and evolution of the universe.
An immediate issue after the introduction of the RS models was the
question of what the form of the field equations on the brane is,
knowing their form in the bulk space. An answer came along in the
elegant work of Shirumizu, Maeda and Sasaki (SMS) \cite{SMS} where
they showed how to, as it were, project the field equations in
the bulk onto the brane in a covariant manner. The subsequent
avalanche of research works have been somewhat overwhelming.
Numerous papers have appeared and tried to remedy the shortcomings
of, generally speaking, the standard model of gravity, including
the late-time acceleration of the universe, the galaxy rotation
curves, the virial mass discrepancy, etc \cite{attempts}.
For a review on the brane word models see \cite{maartens}.

In the SMS method, the bulk geometry is projected onto the brane
using the Gauss-Coddazi equations and Israel junction conditions
\cite{israel}. By means of the Gauss-Coddazi equations, the
intrinsic curvature of the bulk is related to the intrinsic and
extrinsic curvatures of the brane. The Israel junction conditions
then relate the extrinsic curvature of the brane to the
energy-momentum content of the model. In this method, the brane
equations of motion reflect the global geometry of the bulk
space-time through the electric part of the Weyl tensor. However,
there are situations where the use of the Israel junction conditions
become somewhat restrictive. In cases where more than one extra
dimension is involved their use is not well understood. Also, there are
certain matters that are not compatible with these junction
conditions \cite{zahra}. In the latter, one cannot relate the
extrinsic curvature to the matter content of the brane. Such
considerations motivated the idea presented in \cite{maia} and further
developed in \cite{jal1} where the extrinsic curvature from the
Gauss-Codazzi equations is calculated geometrically and the
assumption of  $\mathbb{Z}_2$ symmetry is relaxed. However, the
effects of the global structure of the bulk space is retained by the
field equations. This model has been largely investigated in the
literature \cite{cp}.

Somewhat parallel to the brane-world development, $f(\mc{R})$
theories of gravity appeared on the scene to deal with some of the
shortcomings of the cosmological implications of the standard
general relativity. In such theories, one modifies the
Einstein-Hilbert action by replacing the Ricci scalar $\mc{R}$ with a
generic function $f(\mc{R})$. It is now well-known that such theories are
equivalent to scalar-tensor theories. In fact, both the metric and
Palatini formulations of $f(\mc{R})$ gravity can be rewritten in
terms of certain versions of the Brans-Dicke theory \cite{bransfr}.
These theories with many variations have been successful, for
example, in describing galaxy rotation curves which is generally
believed to be the result of the existence of dark matter. For a review on $f(R)$ gravities see \cite{fR}.

Within the context of a brane-world scenario, $f(\mc{R})$ theories
have been studied recently in \cite{shahab} where the authors follow
the SMS procedure to project the bulk field equations onto the brane
hypersurface. The field equations thus obtained show that the matter
content of the brane-world is not coupled to the bulk $f(\mc{R})$
term.
In this paper, we generalize the work \cite{shahab} to a theory where the use of $\mathbb{Z}_2$ symmetry and the Israel junction condition is relaxed.
As it turns out, since the junction
conditions are not used, the field equations on the brane contain a
new tensor $Q_{\mu\nu}$ which is made out of the extrinsic curvature
and should be calculated through the Gauss-Codazzi equations. The
effect of the non-trivial bulk action is reflected in a new tensor
which is made of the bulk $f(\mc{R})$ function. One must only
project the scalar $f(\mc{R})$ in an obvious manner onto the brane
in order to calculate this tensor. The resulting  field equations
show that the matter content of the brane is non-minimally coupled
to the bulk $f(\mc{R})$. Therefore, in contrast to the work
presented in \cite{shahab}, the non-trivial bulk action can
influence the brane energy-momentum tensor directly.

Recent observations provided by the supernovae legacy survey data
\cite{SNLS} shows that, taking $\omega_{DE}=\mbox{const.}$, the equation
of the state parameter is found to be $\omega_{DE}=-1.04\pm 0.06$. This
data shows that the de-Sitter accelerating phase is a very good
candidate for the late time acceleration of our universe.
In this sense, one may expect that a good cosmological model should predict some matter dominated epoch followed by an accelerated expanding phase. In this paper we consider this issue for the brane world model with general bulk action using the phase space analysis. The method of phase space analysis for cosmological models is very well known and is used for many alternative  theories \cite{tsu,dyna1,dyna2,dynamical}. As it turns out, the model can explain both the matter dominated and accelerated expanding phases, resulting in a universe which starts from an unstable matter dominated phase and eventually falls into the accelerated expanding phase.

The paper is organized as follows: In the next section we construct the model from a general bulk action in $d$ dimensions, relaxing the use of Israel junction condition and the $\mathbb{Z}_2$ symmetry. The model is then restricted to a $5D$ bulk and cosmological solutions are obtained in section \ref{sec4}. In section \ref{sec4.5} a dynamical system analysis is presented which shows that the model can in principle explain a well founded cosmological evolution. Conclusions and final remarks are drawn in the last section.

                            \section{The model}\label{sec2}
We consider a pseudo-Riemannian manifold $V_m$  in which the
background manifold $\bar V_4$ is isometrically embedded by the map
$\mathcal Y: \bar{V}_4\rightarrow V_m$ such that
\begin{align}
\mathcal G_{AB}\mathcal Y^A_{,\mu} \mathcal Y^B_{,\nu}= \bar{g}_{\mu\nu},\qquad \mathcal G_{AB} \mathcal Y^A_{,\mu}\mathcal
N^B_a=0 ,\qquad\mathcal G_{AB}\mathcal N^A_a \mathcal N^B_b=g_{ab}=\pm1.\label{eqa}
\end{align}
where $\mathcal{G}_{AB} (\bar{g}_{\mu\nu})$ is the bulk (brane)
metric ,$\{\mathcal Y ^A\}(\{x^{\mu}\})$ are bulk (brane)
coordinates and $\mathcal N ^A_a$ are $(m-4)$ unit vectors,
orthogonal to the brane. In order to investigate the effects of the
general bulk geometry, we assume a non-trivial bulk action of the
form
\begin{align}
I=\f{1}{2\kappa_m^2}\int\mathrm{d}^mx\sqrt{-\mathcal G}\big[f(\mathcal{R})-2\Lambda\big]+I_M,\label{eq1}
\end{align}
where
\begin{align}
I_M=\int \textmd{d}^mx \sqrt{-\mathcal G}\mathcal L_M,\label{eq1.1}
\end{align}
and $\mathcal{L}_M$ is the matter Lagrangian and $\mathcal{R}$ is
the bulk Ricci scalar. After variation of $I$ with respect to the
bulk metric, $\mathcal{G}_{AB}$, we obtain

\begin{align}
f^\prime(\mathcal{R})\mathcal{R}_{AB}-\frac 1 2 \mathcal{G}_{AB}f(\mathcal{R})+\mathcal{G}_{AB}\Box f^\prime(\mathcal{R})
-\nabla_A\nabla_B f^\prime(\mathcal{R})+\Lambda\mc{G}_{AB}=\kappa_m^2 T_{AB}.\label{eq2}
\end{align}
where prime represents derivative with respect to the argument.
Rearranging the above equation, we obtain the effective Einstein
field equations in the bulk

\begin{align}
G_{AB}\equiv\mathcal{R}_{AB}-\frac 1 2 \mathcal{R}\mathcal{G}_{AB}=S_{AB},\label{eq3}
\end{align}
where
\begin{align}
S_{AB}&=\frac{1}{f^\prime(\mathcal{R})}\bigg[\kappa_m^2 T_{AB}-\Lambda \mathcal{G}_{A B}-\Big(\frac 1 2 \mathcal{R}f^\prime(\mathcal{R})
-\frac 1 2 f(\mathcal{R})
+\Box f^\prime(\mathcal{R})\Big)\mathcal{G}_{AB}+\nabla_A\nabla_B f^\prime(\mathcal{R})\bigg].\label{eq4}
\end{align}
As was mentioned in the Introduction, we use the method introduced
in \cite{maia} to project the bulk geometry onto the brane. In this
sense, perturbations of $\bar{V}_4$ in a small neighborhood of the
brane along a generic transverse direction $
\xi=\xi^a\mc{N}_a~(a=1,2,...,m-4 )$ orthogonal to the brane are
considered as follows
\begin{align}
\mathcal Z^A (x^{\mu},\xi^a)=\mathcal Y^A+(\mathcal L _{\xi} \mathcal Y)^A,\label{eq5}
\end{align}
where $\mathcal L$ represents the Lie derivative and $\xi^a$ denotes
a small parameter along $\mathcal N^A_a$ that parameterizes the
extra noncompact dimensions. The presence of tangent components of
the vector $\xi$ along the submanifold $\bar{V}_4$ can cause some
difficulties because it can induce some undesirable coordinate
gauges, but, as is shown in the theory of geometric perturbations,
it is possible to choose this vector to be orthogonal to the
background \cite{nash}.

Let us now consider the perturbation of the embedding
map along the orthogonal extra dimension $ \mathcal N_a $, giving
local coordinates of the perturbed brane as
\begin{align}
\mathcal Z^A_{,\mu} (x^{\mu},\xi^a)=\mathcal
Y^A_{,\mu}+\xi^a\mathcal N^A_{a,\mu}(x^{\nu}). \label{eq6}
\end {align}
Using the above assumptions, the embedding equations of the
perturbed geometry are given by
\begin{align}
\mathcal G_{\mu\nu}=\mathcal G_{AB}\mathcal Z^A _{~,\mu}\mathcal Z^B _{~,\nu} ,\quad
 \mathcal G_{\mu a}=\mathcal G_{AB} \mathcal Z^A _{~,\mu} \mathcal N^B _{~a} ,\quad
  \mathcal G_{AB}\mathcal N^A_{a} \mathcal N^B _b =g_{ab}.\label{eq7}
\end{align}
Now, use of the embedding equations and the relation
\begin{align}
 \mc{G}^{AB}=\mc{Y}^A_{,\mu}\mc{Y}^B_{,\nu}\bar{g}^{\mu\nu}+\mc{N}^A_a\mc{N}^B_bg^{ab},\label{eq7.5}
\end{align}
allow us to write the metric of the bulk space in the following
form
\begin{align}
\mathcal G_{AB}=\Big(\begin{matrix}
g_{\mu\nu}+A_{\mu c}A^{~c}_\nu & A_{\mu a} \\
A_{\nu b} & g_{ab}
\end{matrix}\Big), \label{eq8}
\end{align}
where
\begin{align}
g_{\mu\nu}=\bar{g}_{\mu\nu}-2\xi^a\bar{K}_{\mu\nu a}+\xi^a\xi^b\bar{g}^{\alpha\beta}\bar{K}_{\mu\alpha a}
\bar{K}_{\nu\beta b},\label{gg}
\end{align}
is the metric of the perturbed brane and $\bar{K}_{\mu\nu a}$ is the
extrinsic curvature of the original brane defined as
\begin{align}
 \bar{K}_{\mu\nu a}=-\mc{G}_{AB}\mc{Y}^A_{,\mu}\mc{N}^B_{a;\nu}.\label{ex}
\end{align}
We also use the notation $A_{\mu c}=\xi^dA_{\mu c d}$ where
\begin{align}
A_{\mu c d}=\mathcal G_{AB} \mathcal N^A_{~d;\mu}\mathcal N^B_c,\label{eq9}
\end{align}
represent the twisting vector fields. The presence of gauge fields
$A_{\mu a}$ tilts the embedded family of sub-manifolds with respect
to the normal vector $\mathcal N^A$. According to our construction,
the original brane is orthogonal to the normal vectors $\mathcal
N^A$. However, from equation (\ref{eq7}), it can be seen that this
is not true for the deformed geometry. Let us introduce
\begin{align}
\mathcal X^A_{,\mu}=\mathcal Z^A_{,\mu}-g ^{ab}\mathcal N^A_a A_{\mu b}.\label{eq10}
\end{align}
One can easily verify that for the set
$\left\{\mc{X}^A_{,\mu},\mc{N}^A_a\right\}$ we have the following
projection relations
\begin{align}
 \mathcal G_{AB}\mathcal X^A_{,\mu} \mathcal X^B_{,\nu}= g_{\mu\nu},\qquad \mathcal G_{AB} \mathcal X^A_{,\mu}\mathcal
N^B_a=0 ,\qquad
\mathcal G_{AB}\mathcal N^A_a \mathcal N^B_b=g_{ab}=\pm1.\label{eqa1}
\end{align}
These define a new family of embedded manifolds whose members are
always orthogonal to $\mathcal N^A$. This new embedding of the local
coordinates can be suitably used for obtaining induced Einstein
field equations on the brane. The extrinsic curvature of a perturbed
brane in the coordinates $\left\{\mc{X}^A_{,\mu},\mc{N}^A_a\right\}$
becomes
\begin{align}
K_{\mu\nu a}&=-\mathcal G_{AB}\mathcal X^A_{,\mu}\mathcal N^B_{a;\nu}\nonumber\\
&=\bar{K}_{\mu\nu a}-\xi^b\bar{K}_{\mu\gamma a}\bar{K}_{\nu\rho b}\bar{g}^{\gamma\rho}\nonumber\\
&=-\frac 1 2 \frac{\partial g_{\mu\nu}}{\partial \xi^a},
\label{eq11}
\end{align}
which is the generalized York relation. In the basis
$\left\{\mc{X}^A_{,\mu},\mc{N}^A_a\right\}$, the Gauss, Codazzi and
Ricci equations are given by
\begin{align}
R_{\alpha\beta\gamma\delta}=2g^{ab}K_{\alpha[\gamma a}K_{\delta]\beta b}+\mathcal R_{ABCD}\mathcal X^A_{,\alpha}\mathcal
X^B_{,\beta}\mathcal X^C_{,\gamma}\mathcal X^D_{,\delta},\label{eq12}
\end{align}
\begin{align}
2K_{\alpha[\gamma c ;\delta]}=2g^{ab}A_{[\gamma a c}K_{\delta]\alpha b}+\mathcal R_{ABCD}\mathcal X^A_{,\alpha}\mathcal
N^B_{,c}\mathcal X^C_{,\gamma}\mathcal X^D_{,\delta},\label{eq13}
\end{align}
and
\begin{align}
2A_{[\mu ab;\nu]}=-2g^{mn}A_{[\mu ma}A_{\nu]nb}-g^{\sigma\rho}K_{[\mu\sigma a}K_{\nu]\rho b}
-\mathcal{R}_{ABCD}\mathcal{X}^C_{,\mu}\mathcal{X}^D_{,\nu}N^B_a N^A_b, \label{eq13.1}
\end{align}
where $\mathcal R_{ABCD}$ and $R_{\alpha\beta\gamma\delta}$ are the
Riemann tensors for the bulk and the brane respectively.
After contracting equation (\ref{eq12}) and using the relation
\begin{align}
 \mc{G}^{AB}=\mc{X}^A_{,\mu}\mc{X}^B_{,\nu}g^{\mu\nu}+\mc{N}^A_a\mc{N}^B_bg^{ab},\label{eq13.2}
\end{align}
one obtains
\begin{align}
R_{\mu\nu}=\big(K_{\mu\alpha c} K_{\nu}^{~ \alpha c}-K_c K_{\mu\nu}^{~~c}\big)+\mathcal R_{AB}\mathcal X^A_{,\mu}\mathcal
X^B_{,\nu}-g ^{ab}\mathcal R_{ABCD}\mathcal N^A_a \mathcal X^B_{,\mu}\mathcal N^C_b \mathcal X^D_{,\nu}.\label{eq14}
\end{align}
So, the Einstein tensor on the brane is in the following form
\begin{align}
G_{\mu\nu}=G_{AB}\mathcal X^A_{,\mu} \mathcal X^B_{,\nu}+Q_{\mu\nu}+g ^{ab}\mathcal R_{AB}\mathcal N^A_a\mathcal
N^B_bg_{\mu\nu}-g ^{ab}\mathcal R_{ABCD}\mathcal N^A_a \mathcal X^B_{,\mu}\mathcal N^C_b \mathcal X^D_{,\nu}
-\f{1}{2}\mc{C}g_{\mu\nu},\label{eq15}
\end{align}
where
\begin{align}
Q_{\mu\nu}=-g^{ab}\big(K^{\gamma}_{\mu a}K_{\gamma\nu b} -K_a K_{\mu\nu b}\big)+\frac 1 2 \big(K_{\alpha\beta a}
K^{\alpha\beta a} -K_a K^a\big)g_{\mu\nu},\label{eq16}
\end{align}
and
\begin{align}
 \mc{C}=g^{ab}g^{cd}\mc{R}_{ABCD}\mc{N}^A_a\mc{N}^B_c\mc{N}^C_b\mc{N}^D_d.\label{eq16.5}
\end{align}
Using the decomposition of the Riemann tensor into the Ricci tensor,
Ricci scalar and Weyl tensor

\begin{align}
\mathcal R_{ABCD}=C_{ABCD}+ \frac {2}{(m-2)} ( \mathcal G_{B[D} \mathcal R_{C]A}-\mathcal G_{A[D} \mathcal R_{C]B} )
+\frac 2{(m-1)(m-2)}\mathcal R(\mathcal G_{A[D}\mathcal G_{C]B}),\label{eq17}
\end{align}
one obtains the generalized Einstein equations on the brane as
\begin{align}
G_{\mu\nu}&=G_{AB} \mathcal X^A_{,\mu} \mathcal X^B_{,\nu}-\mathcal E_{\mu\nu}+Q_{\mu\nu}-\f{1}{2}\mc{C}g_{\mu\nu}+\frac{(m-3)}{(m-2)}g ^{ab}
\mathcal R_{AB}\mathcal N^A_a\mathcal N^B_bg_{\mu\nu}\nonumber\\
&-\frac{(m-4)}{(m-2)}\mathcal R_{AB} \mathcal X^A_{,\mu} \mathcal X^B_{,\nu}+\frac{(m-4)}{(m-1)(m-2)}\mathcal R
g_{\mu\nu},\label{eq18}
\end{align}
where
\begin{align}
\mathcal E_{\mu\nu}=g^{ab}C_{ABCD} \mathcal N^A_a \mathcal X^B_{,\mu}\mathcal N^C_b \mathcal X^D_{,\nu},\label{eq19}
\end{align}
is the electric part of the Weyl tensor $C_{ABCD}$.

In order to obtain the field equations corresponding to action
(\ref{eq1}), we follow Dvali and Shifman \cite{dvali} which have
proposed a mechanism to localize the standard model gauge fields to
the brane. Using this idea,  we may decompose the components of the
energy-momentum tensor of the matter as
\begin{align}
\kappa_4^2\tau_{\mu\nu}=\frac{2\kappa_m^2}{m-2}T_{\mu\nu},\qquad T_{ab}=0,\qquad  T_{\mu a}=0.\label{eq25.1}
\end{align}
However, the geometric part of the generalized energy-momentum
tensor includes the bulk-bulk and the bulk-brane components, and one
must take this into account when writing the Einstein field
equations. We thus define the components of the total energy-momentum
tensor in the basis $\left\{\mc{X}^A_{,\mu},\mc{N}^A_a\right\}$ as
\begin{align}
S_{\mu \nu}=S_{AB}\chi^{A}_{,\mu}\chi^{B}_{,\nu},\qquad S_{\mu a}=S_{AB}\chi^{A}_{,\mu}\mc{N}^{B}_{a},\qquad S_{ ab}=
S_{AB}\mc{N}^{A}_{a}\mc{N}^{B}_{b}.\label{eq22}
\end{align}
After contracting equation (\ref{eq3}), one obtains
\begin{align}
\mathcal R=-\frac 2 {m-2}S,\label{eq23}
\end{align}
where $S=S^A_{~A}$ is the trace of the generalized bulk
energy-momentum tensor. The $\mathcal R_{AB}$ is then given by
\begin{align}
\mathcal R_{AB}=-\frac {1} {m-2}\mathcal G_{AB}S+S_{AB}. \label{eq24}
\end{align}
Substituting $\mathcal R_{AB}$  and $\mathcal R$ in equation
(\ref{eq18}) and using equation (\ref{eq22}), we find

\begin{align}
G_{\mu\nu}=Q_{\mu\nu}-\mathcal E_{\mu\nu}-\f{1}{2}\mc{C}g_{\mu\nu}+\frac{2}{m-2}S_{\mu\nu}
+\frac{m-3}{m-2}g^{ab}S_{ab}g_{\mu\nu}-\frac{(m-4)(m-3)}{(m-2)(m-1)}g_{\mu\nu}S. \label{eq25}
\end{align}

Finally, using equation (\ref{eq25.1}), we obtain the effective
Einstein equations on the brane as
\begin{align}
G_{\mu\nu}-Q_{\mu\nu}+\mathcal E_{\mu\nu}-\f{\kappa_4^2}{f^\prime(\mathcal R)}\tau_{\mu\nu}+\Pi_{\mu\nu}+
\f{1}{2}\mc{C}g_{\mu\nu}=0,\label{eq26}
\end{align}
where $Q_{\mu\nu}$ is defined by equation (\ref{eq16}) and
\begin{align}
\Pi_{\mu\nu}&=Wg_{\mu\nu}-\frac{2}{m-2}\frac{\nabla_{A}\nabla_{B}f^\prime(\mathcal{R})}{f^\prime(\mathcal{R})}\mc{X}^A_{,\mu}
\mc{X}^B_{,\nu}\nonumber\\
&=Wg_{\mu\nu}-\frac{2}{m-2}\frac{\nabla_{\mu}\nabla_{\nu}f^\prime(\mathcal{R})}{f^\prime(\mathcal{R})},
\label{pi}
\end{align}
is a new tensor that reflects the effects of the non-trivial bulk
action on the brane. In the second line of the above equation we
have assumed $\mc{X}^A_{,\mu}=\delta^A_\mu$, so that $\nabla_{\mu}$
is the $\mu= 0,1,2,3$ components of the $m$ dimensional covariant derivative.
We also define
\begin{align}
W=\left(\frac{m^2-5m+10}{2m^2-6m+4}\right)\mathcal{R}&+\left(\frac{m^2-7m+14}{2(m-2)}\right)\left(\frac{2\Lambda-f}
{f^\prime}\right)\nonumber\\&+\left(\frac{m^2-8m+17}{m-2}\right)\frac{~^m\Box
f^\prime}{ f^\prime} +
\frac{m-3}{m-2}\frac{\nabla_{A}\nabla_{B}f^\prime
}{f^\prime}\left(\mc{X}^A_{,\mu} \mc{X}^B_{,\nu} g^{\mu\nu}\right).
\label{W}
\end{align}
These tensors must be evaluated on the brane. We also note that in the limit $f(\mc{R})=\mc{R}$ the tensor $\Pi_{\mu\nu}$ can be converted to the trace of brane energy-momentum tensor, and the field equation \eqref{eq26} reduces to \cite{maia}.

Let us focus our attention on the five dimensional bulk. In the case
of a co-dimension one brane we have $\mc{C}=0$ due to the symmetries
of the Riemann tensor. We also consider the case where
$\mc{X}^A_{,\mu}=\delta^A_\mu$. In this case, equation (\ref{eq13.2})
reduces to
\begin{align}
 \mc{G}_{AB}=g_{AB}+\mc{N}_A\mc{N}_B,\label{eq26.05}
\end{align}
and the field equations (\ref{eq26}) become
\begin{align}
G_{\mu\nu}-Q_{\mu\nu}+\mathcal
E_{\mu\nu}-\kappa_4^2\frac{\tau_{\mu\nu}}{f^\prime(\mathcal{R})}+\Pi_{\mu\nu}=0,\label{eq26.1}
\end{align}
where $\kappa_4^2=\f{2}{3}\kappa_5^2$, and
\begin{align}
\Pi_{\mu\nu}=Wg_{\mu\nu}-L_{\mu\nu},\label{eq30}
\end{align}
\begin{align}
W=\frac {5}{12}\mathcal{R}+L+\frac 2 {3f^\prime}\Big[
{~^5\Box f^\prime}-{f}-2\Lambda\Big],\label{eq31}
\end{align}
\begin{align}
 L_{\mu\nu}=\f{2}{3}\f{\nabla_\mu\nabla_\nu f^\prime}{f^\prime},\qquad L=L^\mu_{~\mu}\label{eq31.1}
\end{align}
with
\begin{align}
Q_{\mu\nu}=(K K_{\mu\nu} - K_{\mu\alpha}K_\nu ^{\alpha})+\frac 1 2
(K_{\alpha\beta}K^{\alpha\beta}-K^2)g_{\mu\nu}, \label{field}
\end{align}
where $K=g^{\mu\nu} K_{\mu\nu}$.  In the case of a constant
curvature bulk, the tensor $Q_{\mu\nu}$ is an independently
conserved quantity as can be seen easily from the Codazzi equation
(\ref{eq13}). However, we are interested in a general bulk geometry,
so we consider  $Q_{\mu\nu}$ as a general tensor which reflects the
bulk effects on the brane equations of motion. By defining the new
tensor
\begin{align}
 M_{\mu\nu}=\kappa_4^2\frac{\tau_{\mu\nu}}{f^\prime(\mathcal{R})}+Q_{\mu\nu}-\mathcal E_{\mu\nu}-\Pi_{\mu\nu}, \label{a}
\end{align}
and using the Bianchi identity, we find that $M_{\mu\nu}$ is
conserved. It is therefore possible to consider $M_{\mu\nu}$ as a
new effective energy-momentum tensor made of the standard conserved
matter energy-momentum tensor and terms reflecting the effects of
the extra dimension and also the non-trivial bulk action. The field
equation on the brane can now be written as
\begin{align}
 G_{\mu\nu}=M_{\mu\nu}. \label{eeq}
\end{align}
                        \section{Cosmological Solutions}\label{sec4}
To study the time evolution of the universe, we take the bulk metric
as
\begin{align}
 ds_B^2=-dt^2+a(t,w)^2\left(\f{dr^2}{1-kr^2}+r^2d\Omega^2\right)+dw^2.\label{lineel}
\end{align}
Taking $\mc{N}^A=\delta^A_5$ and using the equation \eqref{eq26.05}
the brane metric reduces to the usual FRW form
\begin{align}
 ds^2=-dt^2+a(t)^2\left(\f{dr^2}{1-kr^2}+r^2d\Omega^2\right).\label{lineel1}
\end{align}
With this choice, the electric part of the Weyl tensor becomes
\begin{align}
 \mc{E}^0_0=\f{1}{2{a}^2}\left[-{a}\ddot{a}+\dot{a}^2+k+{a}\hat{\hat{a}}-{\hat{a}}^2\right]=-3\mc{E}^i_i\, ,\,\,\,\,\, i=1,2,3,
\end{align}
where dot represents derivative with respect to $t$ and a hat over an arbitrary function $f(r,w)$ is defined as
$\hat{f}(r)=\f{\partial f}{\partial w}\mid_{w=0}$. We can also calculate the extrinsic curvature and
$Q_{\mu\nu}$ by use of equations (\ref{eq11}) and (\ref{field})
respectively, with the result
\begin{align}
K^\mu_{~\nu}=-\left(\f{\hat{a}}{a}\right)\textmd{diag}\left(0,1,1,1\right),\label{kk}
\end{align}
and
\begin{align}
 Q^\mu_{~\nu}=-\left(\f{\hat{a}}{a}\right)^2\textmd{diag}\left(3,1,1,1\right).\label{qq}
\end{align}
By assuming that the matter on the brane has the perfect fluid form
\begin{align}
 \tau^\mu_{~\nu}=\textmd{diag}\left(-\rho,p,p,p\right),\label{ttt}
\end{align}
the conserved tensor $M_{\mu\nu}$ can be written as
\begin{align}
 M^\mu_{~\nu}=\textmd{diag}\left(-\rho_A,p_A,p_A,p_A\right),\label{aaa}
\end{align}
which, using equations (\ref{eq30}), (\ref{a}), (\ref{qq}) and
(\ref{ttt}), result in an effective energy-density and pressure on
the brane
\begin{align}
\rho_A&=\f{\kappa_4^2}{f^\prime}\rho+3\left(\f{\hat{a}}{a}\right)^2+\mc{E}^0_{~0}+W-L^0_{~0},\nonumber\\
p_A&=\f{\kappa_4^2}{f^\prime}p-\left(\f{\hat{a}}{a}\right)^2+\f{1}{3}\mc{E}^0_{~0}-W+L^i_{~i}.\label{aa}
\end{align}
Now, with the aid of equation (\ref{eq31.1}), $L_{\mu\nu}$ becomes
\begin{align}
 L^0_{~0}&=-\f{2}{3}\f{\ddot{f}^\prime}{f^\prime},\nonumber\\
 L^i_{~i}&=-\f{2}{3}\f{1}{a f^\prime}\left(\dot{f^\prime}\dot{a}-\hat{f^\prime}\hat{a}\right),\qquad i=1,2,3.\label{ll1}
\end{align}
Using equation (\ref{eeq}), the Friedmann and the Raychaudhuri equations become
\begin{align}
 H^2&=\f{1}{3}\left[\f{\kappa_4^2}{f^\prime}\rho+3\left(\f{\hat{a}}{a}\right)^2+\mc{E}^0_{~0}+W+\f{2}{3}\f{\ddot{f}^\prime}{f^\prime}
-\f{3k}{a^2}\right],\nonumber\\
\f{\ddot{a}}{a}&=-\f{1}{6}\Bigg[\f{\kappa_4^2}{f^\prime}\left(\rho+3p\right)+2\left(\mc{E}^0_{~0}-W\right)-
\f{2}{3}\f{1}{a f^\prime}\left(3\dot{f^\prime}\dot{a}-3\hat{f^\prime}\hat{a}-\ddot{f^\prime}a\right)\Bigg].\label{eq101}
\end{align}

\subsection{The case $f(\mathcal{R})=f_0e^{\alpha\mathcal{R}}$}
Let us consider first the case $f(\mathcal{R})=f_0e^{\alpha\mathcal{R}}$ where $f_0$ and $\alpha$ are some constants.
Assuming $\tau_{\mu\nu}=0$ and $\Lambda=0$, and using
the trace of the field equation (\ref{eeq})
\begin{align}
L=\f{1}{3}\left(R-\mc{R}+Q+\f{1}{\alpha}\right),\label{L4}
\end{align}
where $R$ is the brane Ricci scalar, $L_{\mu\nu}$ and $W$ take the
following forms
\begin{align}
 L_{\mu\nu}=\f{2}{3}\left(\alpha^2\nabla_\mu\mathcal{R}\nabla_\nu\mathcal{R}
+\alpha\nabla_\mu\nabla_\nu
\mathcal{R}\right),\label{ll}
\end{align}
\begin{align}
 W=\f{1}{12}\left[\f{1}{\alpha}-\mc{R}+4(R+Q)\right].\label{ww}
\end{align}
With these assumptions, equation
({\ref{ww}) reduces to
\begin{align}
 W=-\f{3}{2}\left[\f{\ddot{a}}{a}+H^2-\left(\f{\hat{a}}{a}\right)^2\right]-\f{1}{2}\left(\f{\hat{\hat{a}}}{a}\right)^2
-\f{1}{12\alpha},\label{eq102}
\end{align}
where we have assumed $k=0$.
The solutions of equations ({\ref{eq101}}) are of an inflationary
form, given by
\begin{align}
 a(t)&=c_2 e^{c_1 t},\nonumber\\
\hat{a}(t)&=\beta a(t),\qquad \hat{\hat{a}}(t)=-\left(\beta^2+\f{1}{12\alpha}\right)a(t),\label{eq103}
\end{align}
where $c_1$ and $c_2$ are constants which we choose to be positive
and, $\beta=\beta(\alpha)$ is an arbitrary constant depending only
on $\alpha$. As one can see from the above solutions, derivatives of
the $5D$ scale factor with respect to the extra dimension $w$ is
proportional to the $4D$ scale factor.

We should note
that in our derivation of the field equations for the brane, we did
not use the Codazzi equation. However, the Codazzi equation must be
satisfied for all smooth manifolds. One can easily check that the
above cosmological solutions satisfie the Codazzi equation.
\subsection{The case $f(\mc R)=f_0(\mc R-\mc{R}_0)^{\alpha}$}
Now, we consider a power law form $f(\mc R)=f_0(\mc R-\mc{R}_0)^{\alpha}$, where $f_0$, $\mc R_0$ and $\alpha$ are some constants. Using the trace of the field equations  (\ref{eeq}), one  obtains
\begin{align}
L=\f{1}{3}\left(R+Q-(1-\f{1}{\alpha})\mc{R}-\f{\mc R_0}{\alpha}\right).
\end{align}
Next, using (\ref{eq31}) and (\ref{eq31.1}), one finds
\begin{align}
 W=\f{1}{12}\left[(\f{1}{\alpha}-1)\mc{R}+4(R+Q)-\f{\mc R_0}{\alpha}\right].\label{ww2}
\end{align}
\begin{align}
 L_{\mu\nu}=\f{2}{3}(\alpha-1)\bigg((\alpha-2)\nabla_\mu\mathcal{R}\nabla_\nu\mathcal{R}
+(\mc R-\mc R_0)\nabla_\mu\nabla_\nu
\mathcal{R}\bigg),\label{ll2}
\end{align}
For this specific form of $f(\mc R)$, we again consider the solutions of the equations ({\ref{eq101}}) in the absence of ordinary matter and cosmological constant. For  general $\alpha$ one has a solution
\begin{align}
 a(t)=c_1 t^{1/2},\qquad\hat{a}(t)=c_2a(t),\qquad \hat{\hat{a}}(t)=-\left(c_2^2+\f{1}{6}\mc R_0 \right)a(t),\label{eq1032}
\end{align}
where $c_1$ and $c_2$ are the constants of integration. This form of the scale factor is similar to the radiation dominated epoch.

We have also a power-law solution
\begin{align}
 a(t)=c_1 t^n,\qquad\hat{a}(t)=\sqrt{\f{\mc R_0}{6}}a(t),\qquad \hat{\hat{a}}(t)=-\f{1}{3}\mc R_0 a(t),\label{eq1033}
\end{align}
where
\begin{align}
n=-\f{2\alpha^3-9\alpha^2+4\alpha+6}{\alpha^2+7\alpha-12},
\end{align}
and $c_1$ and $c_2$ are the constants of integration. This solution shows that the above form for the bulk action cannot admit a matter dominated behavior. However, this can be traced to our assumption of the space-time being  empty. In the next section we will see that adding  ordinary matter to the system will produce matter dominated solutions. We have also an inflationary solution as follows
\begin{align}
 a(t)=c_1 e^{nt},\qquad
\hat{a}(t)=c_2 a(t),\qquad \hat{\hat{a}}(t)=\left(2n^2-c_2^2-\f{1}{6}\mc R_0\right) a(t),\label{eq1033-1}
\end{align}
where $n$, $c_1$ and $c_2$ are some constants. The solutions above also satisfy the Codazzi equation and are therefore cosmological solutions of the model.

All the solutions above have the property that derivatives of the scale factor with respect to the extra dimension are proportional to the scale factor itself. In the next section we will use this property to build an autonomous system of equation for the theory.

                              \section{Cosmological Dynamics}\label{sec4.5}
In this section, we investigate the cosmological dynamics of the model introduced in the previous section.
Substitution of expressions for the electric part of the Weyl tensor
and quantity $W$ in equations (\ref{eq101}) results in the
Friedmann and  Raychaudhuri equations for a general $f(\mc{R})$
\begin{align}\label{eq4.5-1}
-\dot{H}-\left(\f{\hat{a}}{a}\right)^2+\f{\hat{\hat{a}}}{a}+\f{1}{4}\f{f}{F}+2\f{\dot{F}H}{F}
-2\f{\hat{F}\hat{a}}{Fa}-\f{\kappa^2}{F}\left(\rho_{rad}+\rho_m\right)=0,
\end{align}
and
\begin{align}\label{eq4.5-2}
\f{4}{3}\dot{H}+\f{2}{3}\f{\hat{\hat{a}}}{a}+\f{4}{3}\left(\f{\hat{a}}{a}\right)^2+\f{2}{3}\f{\ddot{F}}{F}-\f{2}{3}\f{\dot{F}H}{F}
+\f{2}{3}\f{\hat{F}\hat{a}}{Fa}+\f{\kappa^2}{F}\left(\f{4}{3}\rho_{rad}+\rho_m\right)=0,
\end{align}
where we have defined $F(\mc{R})=f^\prime(\mc{R})$. With
definition of the bulk Ricci scalar
\begin{align}\label{eq4.5-3}
\mc{R}=6\left[2H^2+\dot{H}-\left(\f{\hat{a}}{a}\right)^2-\f{\hat{\hat{a}}}{a}\right],
\end{align}
one can write the Friedmann equation as
\begin{align}\label{eq4.5-4}
1=\f{\mc{R}}{12H^2}+\left(\f{\hat{a}}{Ha}\right)^2-\f{1}{8}\f{f}{FH^2}-\f{\dot{F}}{HF}
+\f{\hat{F}\hat{a}}{H^2Fa}+\f{\kappa^2}{2FH^2}\left(\rho_{rad}+\rho_m\right).
\end{align}
In order to study the dynamical evolution of the system, we define the
following dimensionless quantities \cite{tsu}
\begin{align}\label{eq4.5-5}
x_1=\f{1}{H^2}\left[\f{1}{12}\mc{R}+\left(\f{\hat{a}}{a}\right)^2\right],\quad
x_2=-\f{1}{8}\f{f}{H^2F},\quad x_3=-\f{\dot{F}}{HF},\quad
x_4=\f{\hat{F}\hat{a}}{H^2Fa},\quad x_5=\f{\kappa^2\rho_{rad}}{2H^2F},
\end{align}
and
\begin{align}\label{eq4.5-6}
\Omega_m=\f{\kappa^2\rho_m}{2H^2F}=1-\sum_{i=1}^{5}x_i,\qquad
\Omega_{rad}=x_5,\qquad\Omega_{dark}=\sum_{i=1}^{4}x_i.
\end{align}
with physical constraints represented by $\Omega_m,\Omega_{rad}\geq0$. As was
mentioned before, the Codazzi equation must be satisfied for any
solution to the field equations. In our case, it is reduced to
\begin{align}\label{eq4.5-7}
3\f{\dot{\hat{a}}}{a}+\f{\dot{\hat{F}}}{F}=0.
\end{align}
We now assume that the derivatives of the scale factor with
respect to the extra dimension are proportional to the scale factor
itself, that is
\begin{align}\label{eq4.5-8}
\f{\hat{a}}{a}=b,\qquad \f{\hat{\hat{a}}}{\hat{a}}=c,
\end{align}
where $b$ and $c$ are constants which may depend on the detailed functionality of $f(\mc{R})$. This, for example, occurs in the cosmological solution obtained in the previous section.
We also define the dimensionless quantity
\begin{align}\label{eq4.5-8.1}
x_6=-4\f{\mc{R}+6b\left(b+c\right)}{\mc{R}+12b^2}.
\end{align}
The resulting dynamical equations become
\begin{align}\label{eq4.5-9}
\f{\textmd{d}x_1}{\textmd{d}A}=4x_1-\f{x_1x_3}{3m}+ x_1^2x_6,
\end{align}
\begin{align}\label{eq4.5-10}
\f{\textmd{d}x_2}{\textmd{d}A}=4x_2+ x_1x_2x_6+\f{x_1x_3}{2m}+x_2x_3,
\end{align}
\begin{align}\label{eq4.5-11}
\f{\textmd{d}x_3}{\textmd{d}A}=-1+\left(2\beta+1\right)x_1+\f{1}{2}\beta x_1x_6-3x_2-2x_4+x_5+x_3^2+\f{1}{2}x_1x_3x_6,
\end{align}
\begin{align}\label{eq4.5-12}
\f{\textmd{d}x_4}{\textmd{d}A}=-\left(2\beta+6\right) x_1-\f{1}{2}\left(3+\beta\right)x_1x_6+4x_4+ x_1x_4x_6+x_3x_4,
\end{align}
\begin{align}\label{eq4.5-13}
\f{\textmd{d}x_5}{\textmd{d}A}=x_1x_5x_6+x_3x_5,
\end{align}
and
\begin{align}
\f{\textmd{d}x_6}{\textmd{d}A}=\f{1}{3m}x_3\left(x_6+4\right),\label{eq4.5-13.1}
\end{align}
where $A=\ln a$ and we have defined
\begin{align}\label{eq4.5-14}
\beta=\f{3c}{b-c}.
\end{align}
We also define the quantity
\begin{align}\label{eq4.5-14.5}
m=\left(\f{\mc{R}}{3}+4b^2\right)\f{F^\prime}{F},
\end{align}
which depends on the bulk Ricci scalar. From equation
(\ref{eq4.5-8.1}) we see that the Ricci scalar can be represented as
a function of the quantity $x_6$, so one may consider the quantity
$m$ as a function of $x_6$.

Let us define the effective equation of state $p_{tot}=\omega_{eff}\rho_{tot}$, where
\begin{align}\label{eq4.5-17}
\omega_{eff}=-1-\f{2}{3}\f{\dot{H}}{H^2}=\f{1}{3}\left(1+ x_1x_6\right).
\end{align}
With these definitions, $\omega_{eff}=0$ gives the behavior
$a\propto t^{\f{2}{3}}$ which is the standard behavior of the matter
dominated epoch. Also $\omega_{eff}=-1$ corresponds to the de Sitter
epoch.

In order to find the dynamical behavior of the above cosmological
model, one must obtain the critical points of the dynamical system
(\ref{eq4.5-9})-(\ref{eq4.5-13.1}) followed by expanding the system
near the critical point. The resulting equations can be written as
\begin{align}\label{eq4.5-19}
\mathbf{X^\prime}=\Xi\mathbf{X},
\end{align}
where $\textbf{X}$ is a column vector made out of $x_i$'s, and the
prime represents derivative with respect to  parameter $A=\ln a$ while
$\Xi$ is a matrix obtained from linearizing the system
(\ref{eq4.5-9})-(\ref{eq4.5-13.1}). The eigenvalues of this matrix at
each critical point determine the stability of that point. If all
the real parts of the eigenvalues are negative, then the point is
stable. The appearance of positive eigenvalues makes the point a
saddle point. If, however, all real parts of the eigenvalues are
positive, we have an unstable point. In the evolution of the
universe, one needs a saddle or an unstable point which would correspond
to the matter epoch. Because of the instability of the this phase,
it can fall into a stable point which would then correspond to an
accelerating phase.

The critical points of the dynamical system
(\ref{eq4.5-9})-(\ref{eq4.5-13.1}) are discussed in what follows:
\subsection{Radiation epoch}
\begin{align}\label{eq4.5-19.01}
P_1:&\left(x_1,x_2,x_3,x_4,x_5,x_6\right)=\left(0,0,0,0,1,x\right),\nonumber\\
&\omega_{eff}=\f{1}{3},\qquad\Omega_m=0,\qquad\Omega_{rad}=1,\qquad\Omega_{dark}=0,
\end{align}
where $x$ is arbitrary. This is the standard radiation dominated epoch, with $a\propto
t^{\f{1}{2}}$, using  equation (\ref{eq4.5-17}). The eigenvalues of
this point are
\begin{align}\label{eq4.5-19.1}
0,\quad1,\quad-1,\quad4,\quad4,\quad4,
\end{align}
making this point is a saddle point as discussed above.

\subsection{Kinetic epoch}
We have three critical points which correspond to the kinetic epoch
of the universe as follows
\begin{align}\label{eq4.5-20}
P_2:&\left(x_1,x_2,x_3,x_4,x_5,x_6\right)=\left(0,5,-4,0,0,-4\right),\nonumber\\
&\omega_{eff}=\f{1}{3},\qquad\Omega_m=0,\qquad\Omega_{rad}=0,\qquad\Omega_{dark}=1.
\end{align}
\begin{align}\label{eq4.5-21}
P_3:&\left(x_1,x_2,x_3,x_4,x_5,x_6\right)=\left(0,0,1,0,0,-4\right),\nonumber\\
&\omega_{eff}=\f{1}{3},\qquad\Omega_m=0,\qquad\Omega_{rad}=0,\qquad\Omega_{dark}=1,
\end{align}
and
\begin{align}\label{eq4.5-22}
P_4:&\left(x_1,x_2,x_3,x_4,x_5,x_6\right)=\left(0,0,-1,0,0,-4\right),\nonumber\\
&\omega_{eff}=\f{1}{3},\qquad\Omega_m=2,\qquad\Omega_{rad}=0,\qquad\Omega_{dark}=-1.
\end{align}
The eigenvalues to these points are
\begin{align}\label{eq4.5-22.1}
P_2:&0, -5,-4,-3,\f{4(1+3m)}{3m},-\f{4}{3m},\\
P_3:& 1,2,\f{12m-1}{3m},5,5,\f{1}{3m},\\
P_4:& -2,-1,\f{12m+1}{3m},3,3,-\f{1}{3m}.
\end{align}
As can be seen, point $P_4$ is a saddle point. The point $P_2$ is either stable or saddle, and the point $P_3$ is either saddle or unstable depending on the sign of $m$.

\subsection{de Sitter epoch}
This epoch includes four critical points corresponding to the
accelerating phase of the universe. The one which can be used to describe the late time acceleration
of the universe is the stable point. However, unstable accelerated fixed points can be used to describe the inflation epoch to ensure the end of the inflationary accelerating phase. The critical points
corresponding to this epoch are
\begin{align}\label{eq4.5-23}
&P_5:\left(x_1,x_2,x_3,x_4,x_5,x_6\right)=\left({\frac {3m}{1+3\,m}}\, ,{\frac {-9m}{2\left(1+3m\right)^2}}\, ,
{\frac {12m}{1+3\,m}}\, ,0,\,-\f{144m^2+21m-2}{2\left(1+3m\right)^2},-4\right)\nonumber\\
&\omega_{eff}={\frac {1-9\,m}{3+9\,m}},\qquad\Omega_m=\frac {6  m\left(1+6m \right)}{\left(1+3m\right)^2},\quad\Omega_{rad}=-\f{144m^2+21m-2}{2\left(1+3m\right)^2},\quad
\Omega_{dark}=\frac { 3 m\left( 30\,m+7 \right)}{2\left(1+3m\right)^2}.
\end{align}
For $m<-\f{1}{3}$ this point corresponds to the phantom epoch, and it
is also compatible with the condition $\Omega_m\geq0$, but it gives $\Omega_{rad}<0$. The
corresponding eigenvalues  are then given by
\begin{align}\label{eq4.5-24.1}
4,\quad 4,\quad\f{4}{1+3m},\quad\f{\alpha_1+4m\gamma}{2\gamma(1+3m)},\quad-\f{\alpha_1-8m\gamma\pm i \sqrt{3}\alpha_2}{4(1+3m)\gamma},
\end{align}
where we have defined
\begin{align}\label{eq4.5-24.2}
\alpha_{1,2}=\left(-2240m^3-1572m^2-156m+16+2\sqrt{M}\right)^{\f{2}{3}}\pm\left(208m^2+14m-6\right),
\end{align}
and
\begin{align}
\gamma&=\left(-2240m^3-1572m^2-156m+16+2\sqrt{M}\right)^{\f{1}{3}},
\end{align}
with
\begin{align}
M=-995328{m}^{6}+1306368{m}^{5}+956628{m}^{4}+130218{m}^{3}-11226{m}^{2}-1626m+118.
\end{align}
The critical point $P_5$ is thus a saddle point and together with the result  $\Omega_{rad}<0$ cannot serve as a final accelerating phase for the universe. The next point
is
\begin{align}\label{eq4.5-26}
P_6:&\left(x_1,x_2,x_3,x_4,x_5,x_6\right)=\left(1,0,0,0,0,-4\right),\nonumber\\
&\omega_{eff}=-1,\qquad\Omega_m=0,\qquad\Omega_{rad}=0,\qquad\Omega_{dark}=1,
\end{align}
which corresponds to the standard de Sitter epoch. The eigenvalues
corresponding to this point are
\begin{align}\label{eq4.5-27}
0,\quad 0,\quad -4, \quad \delta_1-2,\quad -\f{1}{2}\delta_1-2\pm i \f{\sqrt {3}}{2} \delta_2,
\end{align}
where
\begin{align}\label{eq4.5-28}
\delta_{1,2}={\f{\gamma}{6m}}\pm{\f{24\,m-11}{3\gamma}},\quad
\gamma=m^{\f{2}{3}}\left( 2\sqrt {{\f{M}{m}}}-252 \right)^{\f{1}{3}},
\end{align}
and
\begin{align}\label{eq4.5-30}
M=2662-1548m+38016m^2-27648m^3.
\end{align}
So for $-4<\delta_1<2$ we have a stable de Sitter epoch which, when written in terms of $m$, leads to
$m<-\f{5}{48}$, $0<m<\f{11}{24}$ or $m>1.38$. The point $P_6$ corresponds to
an accelerating epoch much in the same way as the cosmological
constant. As was mentioned in the Introduction, the observational
data predict that the accelerating behavior of the late time
universe is very close to that of the cosmological constant. Our
model is therefore compatible with such predictions.  Finally we
have two other critical points which can be written as
\begin{align}\label{eq4.5-31}
P_{7,8}:\left(x_1,x_2,x_3,x_4,x_5,x_6\right)=\left(\f{-2}{3}x(1+3m),x,4m\big[2x(1+3m)+3\big],0,0,-4\right),
\end{align}
\begin{align}\label{eq4.5-31.1}
\omega_{eff}&=\f{1}{3}\left[\f{8}{3}x(1+3m)+1\right],\nonumber\\
\Omega_m&=1-\Omega_{dark},\quad\Omega_{rad}=0,\quad\Omega_{dark}=\f{1}{3}x+6m\bigg[x(1+4m)+2\bigg],
\end{align}
where
\begin{align}\label{eq4.5-32}
x={\frac {1}{64}}\,{\frac {11-42\,m-720\,{m}^{2}-1728\,{m}^{3}\pm\sqrt
{20736\,{m}^{4}-15552\,{m}^{3}-9468\,{m}^{2}-540\,m+121}}{m \left( 1+
12\,m+45\,{m}^{2}+54\,{m}^{3} \right) }}.
\end{align}
The point $P_7$ is interesting since it represents a phantom
epoch for $m<-\f{1}{3}$ or $-\f{1}{6}<m<0$. In figure \ref{p7} we
have shown the behavior of $\omega_{eff}$ as a function of $m$. However,
we must impose the physical condition $\Omega_m\geq 0$ to this point
which leads to $-0.21<m<0$. Figure \ref{E} shows four
different eigenvalues of the critical point $P_7$ as a function of $m$. As we
can see from the figure, only in the case $-\f{1}{6}<m<0$ all
eigenvalues are negative. So the critical point $P_7$ is stable if
$-\f{1}{6}<m<0$. This is interesting since
such a range is also physical and can be used for the stable
accelerated expansion phase of the universe.
\begin{figure}
\centering
 \includegraphics[scale=0.9]{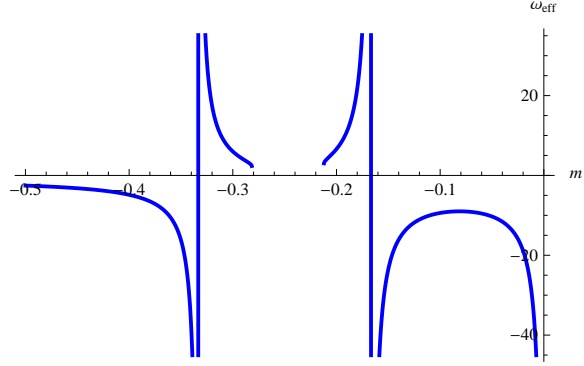}
\caption{Plot of $\omega_{eff}$ as a function of $m$ for $P_7$. The physical range for the stable accelerating epoch
is $-\f{1}{6}<m<0$.}\label{p7}
\end{figure}
\begin{figure}
\centering
 \includegraphics[scale=0.67]{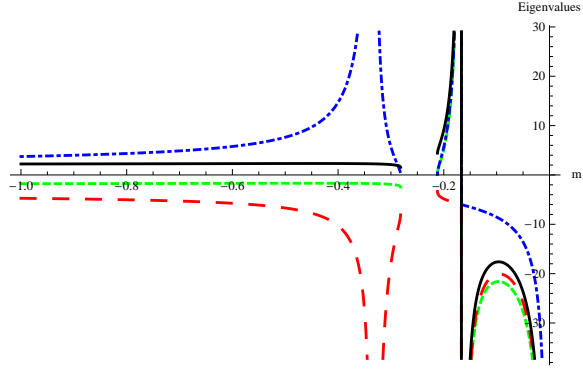}
\caption{Plot of the eigenvalues of the critical point $P_7$. As can be seen, for the range $0<m<-\f{1}{6}$, all the eigenvalues are
negative.}\label{E}
\end{figure}
The critical point $P_8$ has an accelerating de Sitter phase for $|m|\gg 1$
\begin{figure}
\centering
 \includegraphics[scale=0.9]{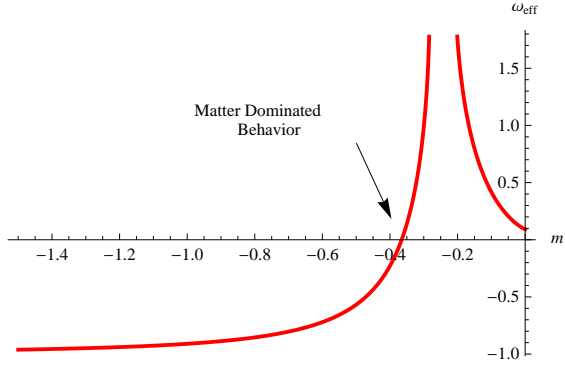}
\caption{Plot of $\omega_{eff}$ as a function of $m$ for the critical point
$P_8$. The asymptotic value for $\omega_{eff}$ in the limit $m\rightarrow\infty$
represents the de Sitter phase with $\omega_{eff}=-1$. The matter dominated phase is represented by $m=-0.36$ as shown in the figure.}\label{p8}
\end{figure}
\begin{align}\label{eq4.5-32.1}
\omega_{eff}=-1,\qquad \Omega_m=-\f{1}{4m}.
\end{align}
The eigenvalues for the corresponding point in this limit are
\begin{align}\label{eq4.5-33}
\f{3}{4m},\quad -2+\f{4}{3m}, \quad -4+\f{1}{6m},\quad \f{3}{4m},\quad \f{3}{4m},\quad0.
\end{align}
The point $P_8$ is thus stable provided  $m<0$. In figure \ref{p8} we have  plotted $\omega_{eff}$ as a function of $m$ for the point $P_8$. As can be seen, $\omega_{eff}$ reaches its minimum value at minus infinity which  corresponds to the de Sitter accelerating phase. Figure  \ref{p8} shows another interesting feature of the critical point $P_8$. The value $m=-0.36$ corresponds
to  $\omega_{eff}=0$ which has the behavior of a matter dominated epoch, $a(t)\propto t^{\f{2}{3}}$,  with
\begin{align}\label{eq4.5-34}
\Omega_m=0.14,\qquad\Omega_{rad}=0,\qquad\Omega_{dark}=0.86.
\end{align}
At the fix point, we have from \eqref{eq4.5-14.5}
\begin{align}
 f(\mc{R})=f_0\left(\mc{R}+2b^2\right)^{\f{27}{25}}.
\end{align}
The eigenvalues corresponding to this point are
\begin{align}\label{eq4.5-35}
-0.3,\quad 3, \quad -4.6,\quad -3.5,\quad -0.3,\quad-0.3.
\end{align}
This point is therefore a saddle point and represents the matter dominated era of the universe.

The discussion in this section suggests that the generalized model proposed here has a dominant accelerated expansion behavior. This is desirable since the brane-world models are well-known for the production of late time accelerated behavior
of the universe. However as we saw in this section, the effect of the non-trivial bulk action results in  changes in the behavior of the scale factor and the dominant contribution is that of dark energy.

                                                \section{Conclusions and final remarks}\label{sec6}
\begin{figure}
\centering
 \includegraphics[scale=0.85]{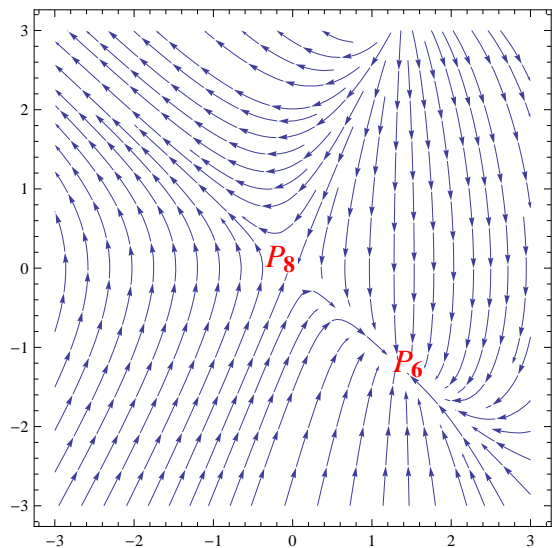}
\caption{Stream plot of the point $P_8$ and $P_6$ which correspond to the unstable matter dominated phase and  stable accelerating phase respectively.}
\label{dyna}
\end{figure}

In this work we have considered a brane-world scenario in the context of
$f(\mc{R})$ gravity. We showed that the change in the dynamics of the bulk space-time leads to changes in the dynamics of the brane space-time. The change to the brane equation of motion is however important, for we have the non-minimal coupling of the brane energy-momentum tensor to the bulk $f(\mc{R})$, which is similar to 4-dimensional $f(R)$ gravity models. This would not have been the case had we used Israel junction conditions to project the bulk geometry to the brane \cite{shahab}.

The cosmological solution found with the exponential form of $f(\mc{R})$ exhibits
an inflationary behavior. The power law assumption for the form of $f(\mc{R})$ was also considered. The main point of this ansatz is that if we consider the vacuum space-time one cannot obtain  matter dominated solutions for the theory. However, this difficulty can be resolved if one adds an ordinary matter to the theory. On the other hand, the radiation dominated solution always exists. It is worth noting that all the solutions obtained in this paper have the property that the derivative of the scale factor with respect to the extra dimension are always proportional to the scale factor itself. Of course solutions with general extra dimensional derivatives may also exist.

In order to describe the cosmological evolution of the universe, one must consider the general form of $f(\mc{R})$. This was done by studying the phase space of the theory. The interesting result was that there is an unstable critical point $P_8$, which can be considered as a
matter dominated phase of the universe. This point corresponds to a power law form for $f(\mc{R})$ and can fall into a stable accelerating critical point $P_6$ which leads to the value $m=-0.36$.
Figure \eqref{dyna} shows a stream plot of these two points.
We also note that with the assumption (\ref{eq4.5-8}), we obtained a new dynamical variable which contains derivative of $f(\mc{R})$ with respect to the extra dimension. This dynamical variable is a new feature of the model and is responsible for the change of the dynamics of the system relative to what one gets in the corresponding $4D$ scenario \cite{tsu}. The complete analysis of the system would be more complicated if we relaxed the assumption (\ref{eq4.5-8}) which reflects the full behavior of the non-trivial dynamics of the bulk space-time.

                              
\end{document}